# Fixed points on band structures of non-Hermitian models: Extended states in the bandgap and ideal superluminal tunneling

Amir M. Jazayeri

*The College of Optics and Photonics (CREOL), University of Central Florida, Orlando, FL, USA*

*amir@ucf.edu*

Space-time reflection symmetry (**PT** symmetry) in non-Hermitian electronic models has drawn much attention over the past decade mainly because it guarantees that the band structures calculated under open boundary conditions be the same as those calculated under periodic boundary conditions. **PT** symmetry in electromagnetic (EM) models, which are usually borrowed from electronic models, has also been of immense interest mainly because it leads to 'exceptional' parameter values below which non-Hermitian operators have real eigenvalues, although **PT** symmetry is not the sole symmetry which allows such exceptional parameter values. In this article, we examine one-dimensional **PT**-symmetric non-Hermitian EM models to introduce novel concepts and phenomena. We introduce the band-structure concept of 'fixed points', which leads to 'bidirectional' reflection zeros in the corresponding finite structures, contrary to a common belief about the EM structures with **PT** symmetry (and without **P** and **T** symmetries). Some of the fixed points manifest themselves as what we name 'extended states in the bandgap' on the band structure while some other fixed points are the 'turning points' of the band structure. The extended states in the band gap are in fact the dual of the well-known 'bound states in the continuum' while the turning points allow us to observe 'ideal' superluminal tunneling in the corresponding finite structures. By 'ideal' superluminal tunneling we mean the case where not only the transmission coefficient has an almost uniform phase over a broad bandwidth but also the magnitudes of the transmission and reflection coefficients are almost equal to unity and zero, respectively, over the bandwidth.

## I. INTRODUCTION

Spinless electronic systems without **T** (viz., time-reversal) symmetry are allowed to support protected edge states [1-3], whose electromagnetic (EM) counterparts can be realized in structures made of materials with non-reciprocal EM response [4]. Spinful electronic systems with **T** symmetry are also allowed to support protected edge states [5-7], whose EM counterparts have been realized in structures made of ring resonators [8] and bi-anisotropic elements [9], where the role of spin is played by the EM wave's propagation direction and polarization, respectively.

Strictly speaking, parity inversion (viz., **P**) is the reversal of either one spatial coordinate axis or all the three spatial coordinate axes. Sometimes the reversal of two spatial coordinate axes is also called parity inversion, although it is in fact a π-rotation. Unlike **T**, which is always non-Hermitian, always anti-unitary (leading to **T**i**T**$^{-1}$=-i), and sometimes non-involutory (viz., **T**$^{-1}$ and **T** are sometimes different), **P** (as well as the π-rotation operator) is always Hermitian, unitary, and involutory.

The composition of **P** and **T** symmetries is especially important in non-Hermitian systems. For instance, the Hermitian Su-Schrieffer-Heeger (SSH) model [10] has two well-known non-Hermitian versions; one which respects **PT** symmetry [11,12], and the other which does not [11,13]. Non-Hermiticity in the latter, which does *not* respect **PT** symmetry, completely changes the topological character of the original SSH model [11,13] and leads to a fundamental difference between the energy spectrum calculated under periodic boundary conditions and the one calculated under open boundary conditions [11,13] *however large* the length of the open sample is chosen.

Another well-known consequence of **PT** symmetry in non-Hermitian systems is that all the eigenvalues of a **PT**-symmetric non-Hermitian Hamiltonian can be real in a parameter range [14]. The parameter value at which some (or all) of the eigenvalues start to become complex is called an exceptional point. However, we should emphasize that **PT** symmetry is not the sole symmetry which allows non-Hermitian Hamiltonians to have real eigenvalues. For instance, the non-Hermitian version of the SSH model which does not respect **PT** symmetry *does* have real energy spectra in certain parameter ranges, but, due to lack of **PT** symmetry, its energy spectrum and exceptional points depend on the choice of boundary conditions.

The **PT**-symmetric non-Hermitian EM structures reported in the literature are usually borrowed from tight-binding electronic models, and realized by parallel waveguides weakly coupled to each other, where the role of energy in the electronic models is played *mathematically* by the EM momentum along the waveguides [15-18]. As the correspondence is solely mathematical, a certain *physical* property of such an EM structure does *not* correspond to the *same physical* property of its corresponding electronic model. In particular, the EM band structure, which shows the relationship between the photon's energy and momentum along the waveguides (or quasimomentum perpendicular to the waveguides), does *not* correspond to the electronic band structure, which shows the relationship between the electron's energy and quasimomentum.

The **PT**-symmetric non-Hermitian EM models to be examined in this article do not have any *tight-binding* electronic counterparts. Furthermore, the regions of EM gain and loss in the to-be-examined EM structures are so thin in comparison with the photon's wavelength that we approximate them by negative and positive surface conductivities, respectively. We can consider the to-be-examined models as 'generalized' Kronig-Penney models. Unlike the delta functions in the potential profile of the well-known Kronig-Penney electronic model [19], the delta functions in the permittivity profiles of the to-be-examined EM models are not real-valued and are not necessarily equally spaced. The purpose of this work is to introduce certain novel concepts and phenomena, but it is worth noting that 'negative' surface conductivity can, for instance, be realized by using an optically pumped monolayer of graphene and transition-metal dichalcogenides when the EM frequency is in the THz [20,21] and optical range [22], respectively.

## II. PT-SYMMETRIC NON-HERMITIAN EM MODELS

### A. Revisiting a well-known model and phenomenon

A well-known **PT**-symmetric EM structure, which was first examined in [23], is based on a one-dimensional complex refractive index profile $n_b+n_c\cos(2\pi x/\Lambda)+in_s\sin(2\pi x/\Lambda)$, where $n_b$, $n_c$, and $n_s$ are positive numbers at the excitation frequency, and $n_c, n_s \ll n_b$. Let us consider a finite-length structure based on this refractive index profile within $x=0$ and $x=w\equiv M\Lambda$ in a background of a refractive index $n_b$, where M is a positive integer. It is evident that such a finite structure is asymmetric when $n_s$ is non-zero, because the left side of the structure starts with a region imposing EM loss while the right side starts with a region yielding EM gain. Let us write the reflected and transmitted electric field phasors as

$$\hat{y}E_0 R_L \exp(-in_b k_0 x) \text{ and } \hat{y}E_0 T_{LR}\exp(in_b k_0(x-w)), \tag{1}$$

respectively, when the structure is illuminated from the left (viz., at $x=0$) by an incident plane wave with the electric field $\hat{y}E_0\exp(in_b k_0 x)$, and as

$$\hat{y}E_0 R_R \exp(in_b k_0(x-w)) \text{ and } \hat{y}E_0 T_{RL}\exp(-in_b k_0 x), \tag{2}$$

respectively, when the structure is illuminated from the right (viz., at $x=w$) by an incident plane wave with the electric field $\hat{y}E_0\exp(-in_b k_0(x-w))$. We have denoted the wavenumber in free space by $k_0$, which reads $\omega/c$ in terms of the speed of light in free space (c) and the excitation angular frequency ($\omega$).

For $n_s=0$, we know that at the Bragg frequency [viz., when $\omega=\omega_B\equiv\pi c/(n_b\Lambda)$], the structure is a symmetric Bragg mirror, and the reflection coefficients $R_L$ and $R_R$ are equal, and approach unity when M→∞. However, for $n_s=n_c$, the authors of [23] observed that at the Bragg frequency, $R_L$ is zero while $R_R$ is an unbounded and increasing function of M. The fact that for any non-zero ns, the reflection coefficients $R_L$ and $R_R$ are unequal was termed a "non-reciprocal" behavior in Ref. [23], but it should be noted that the structure is in fact reciprocal (i.e., obeys Lorentz reciprocity), and the transmission coefficients $T_{LR}$ and $T_{RL}$ are always equal. These equal coefficients will hereafter be denoted by $T$.

The structure was later re-examined in [24]. For $n_s=n_c$, the authors of [24] observed that at the Bragg frequency, the magnitude of $T$ is almost unity, and its phase can be approximated by $n_b w\omega/c$ in terms of $\omega$. Therefore, for $n_s=n_c$, the structure is almost 'invisible' (viz., acts as a medium of the refractive index $n_b$) when it is illuminated from the left by an incident wave packet with a frequency content around the Bragg frequency.

To the best of our knowledge, the structure has only been examined from the scattering-matrix viewpoint in the literature. Therefore, it is worth looking into the band diagram of its underlying periodic refractive index profile [viz., $n_b+n_c\cos(2\pi x/\Lambda)+in_s\sin(2\pi x/\Lambda)$, where $n_c, n_s \ll n_b$] so we can later contrast its features with the band diagrams of the EM models to be introduced in this article. However, instead of working with the 'refractive index' $n(x)$ defined above, we work with the 'relative permittivity' $\varepsilon(x)$ defined as $\varepsilon_b + \varepsilon_c \cos(2\pi x/\Lambda) + i\varepsilon_s \sin(2\pi x/\Lambda)$, where $\varepsilon_b$, $\varepsilon_c$, and $\varepsilon_s$ are positive numbers. The relative permittivity $n^2(x)$ corresponding to the refractive index $n(x)$ is a special case of $\varepsilon(x)$ when $\varepsilon_c, \varepsilon_s \ll \varepsilon_b$; in this case, the coefficients of $\varepsilon(x)$ can be written as $\varepsilon_b = n_b^2$, $\varepsilon_c = 2n_b n_c$, and $\varepsilon_s = 2n_b n_s$ in terms of the coefficients of $n(x)$. However, by using the standard Fourier expansions of the EM fields [25], we can calculate the band structure of $\varepsilon(x)$ without necessarily assuming that $\varepsilon_c, \varepsilon_s \ll \varepsilon_b$.

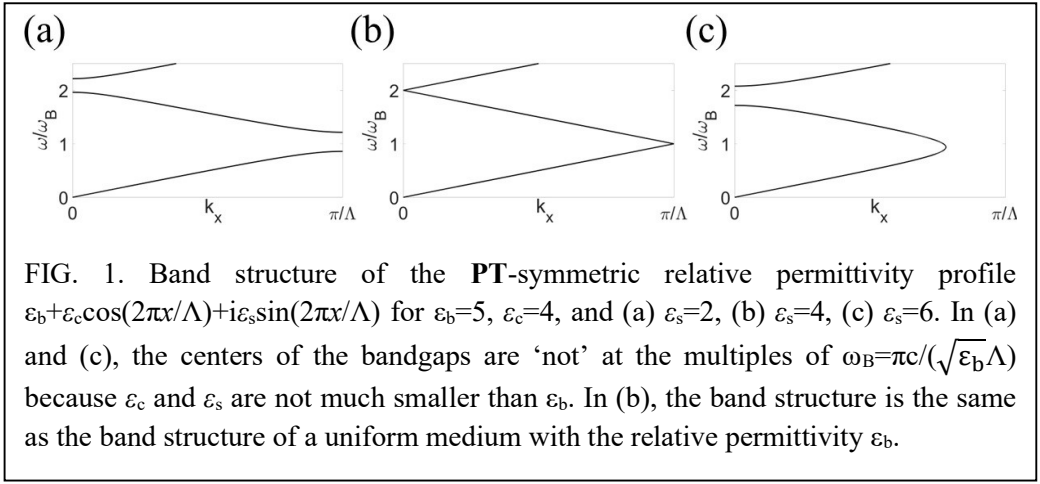

FIG. 1. Band structure of the **PT**-symmetric relative permittivity profile $\varepsilon_b + \varepsilon_c \cos(2\pi x/\Lambda) + i\varepsilon_s \sin(2\pi x/\Lambda)$ for $\varepsilon_b = 5$, $\varepsilon_c = 4$, and (a) $\varepsilon_s = 2$, (b) $\varepsilon_s = 4$, (c) $\varepsilon_s = 6$. In (a) and (c), the centers of the bandgaps are 'not' at the multiples of $\omega_B = \pi c/(\sqrt{\varepsilon_b}\Lambda)$ because $\varepsilon_c$ and $\varepsilon_s$ are not much smaller than $\varepsilon_b$. In (b), the band structure is the same as the band structure of a uniform medium with the relative permittivity $\varepsilon_b$.

We only plot the band structure for $0 < k_x < \pi/\Lambda$ because the band structure is symmetric with respect to $k_x = 0$. As is seen in Fig. 1(a), for $\varepsilon_s < \varepsilon_c$, the band structure has bandgaps. The frequencies located in each bandgap correspond to complex quasimomentums whose real parts are identical (either zero or $\pi/\Lambda$). The bandgap-center frequencies, at which the imaginary parts of the complex quasimomentums are maximized, happen at the multiples of the Bragg frequency $\omega_B = \pi c/(\sqrt{\varepsilon_b}\Lambda)$ if and only if $\varepsilon_c, \varepsilon_s \ll \varepsilon_b$, because in this case, the sine and cosine parts of $\varepsilon(x)$ can be considered as small perturbations added to the constant $\varepsilon_b$.

As is seen in Fig. 1(b), the bandgaps all narrow when $\varepsilon_s$ increases, and they all close when $\varepsilon_s$ becomes equal to $\varepsilon_c$. This confirms that $\varepsilon_s = \varepsilon_c$ is in fact an exceptional parameter value. The band structure for $\varepsilon_s = \varepsilon_c$ is the same as the band structure for a uniform medium with the relative permittivity $\varepsilon_b$. This simply explains the 'invisibility' reported in [24] for a finite structure with $\varepsilon_s$ equal to $\varepsilon_c$ and excitation frequencies around $\omega_B$. It is worth noting that the band crossings for $\varepsilon_s = \varepsilon_c$ all happen at the multiples of $\omega_B$ whether $\varepsilon_c$ is small or not.

As is seen in Fig. 1(c), bandgaps appear again when $\varepsilon_s > \varepsilon_c$. These bandgaps are in fact thanks to a non-zero $\varepsilon_s$ and remain open even if $\varepsilon_c = 0$. However, a more obvious signature of the spontaneously broken **PT** symmetry for $\varepsilon_s > \varepsilon_c$ is the emergence of k-jumps. In the range between any two consecutive bandgaps, the Bloch wave propagating parallel to $+\hat{x}$ (or $-\hat{x}$) has a positive (or negative) k-jump from $k_g$ to $2\pi/\Lambda-k_g$ (or from $2\pi/\Lambda-k_g$ to $k_g$) at some frequency, where $0<k_g<\pi/\Lambda$.

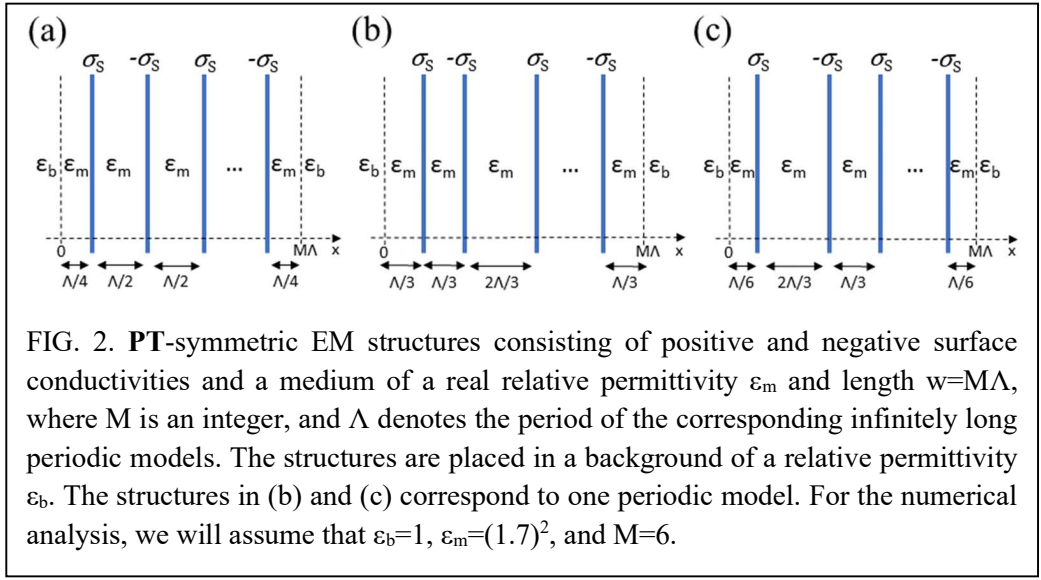

FIG. 2. **PT**-symmetric EM structures consisting of positive and negative surface conductivities and a medium of a real relative permittivity $\varepsilon_m$ and length w=M$\Lambda$, where M is an integer, and $\Lambda$ denotes the period of the corresponding infinitely long periodic models. The structures are placed in a background of a relative permittivity $\varepsilon_b$. The structures in (b) and (c) correspond to one periodic model. For the numerical analysis, we will assume that $\varepsilon_b=1$, $\varepsilon_m=(1.7)^2$, and M=6.

### B. Novel models, concepts, and phenomena

Let us now focus our attention on the **PT**-symmetric structures depicted in Figs. 2, each of which consists of surface conductivities $\sigma_S$ and -$\sigma_S$ and a medium of a relative permittivity $\varepsilon_m$ and length w=M$\Lambda$, where the values of $\sigma_S$ and $\varepsilon_m$ are real and positive at the excitation angular frequency $\omega$, M is an integer, and $\Lambda$ denotes the period of the corresponding infinitely long periodic models. The structures are placed in a background of a positive relative permittivity $\varepsilon_b$. A surface conductivity $\sigma_S$ at a position $x$=a is in fact a volume conductivity $\sigma(x)$ in a region so thin in comparison with the photon's wavelength that we approximate $\sigma(x)$ by $\sigma_S\delta(x$-a), where $\delta(x)$ denotes the Dirac delta function. From an EM viewpoint in the frequency domain, a conductivity $\sigma$ is in fact equivalent to a complex relative permittivity $\varepsilon$ equal to $i\sigma/(\varepsilon_0\omega)$, where the time variation $e^{-i\omega t}$ has been assumed. Positive conductivity (or, equivalently, a positive imaginary part for permittivity) imposes EM loss while negative conductivity (or, equivalently, a negative imaginary part for permittivity) yields EM gain.

Given the geometries of the structures depicted in Fig. 2, we use the standard transfer-matrix method [26] to analyze them at any excitation frequency $\omega$. A transfer matrix is a matrix which relates the EM fields $E_y$ and $H_z$ at a position $x_1$ to the EM fields at another position $x_2$. The scattering-matrix elements [viz., $R_L$, $R_R$, and $T$, defined in Eqs. (1) and (2)] for each of the structures depicted in Fig. 2 can be built based on the transfer matrix relating the EM fields at $x$=0 to the EM fields at $x$=w≡M$\Lambda$. We can also use the transfer-matrix method to derive the band structures of the corresponding infinitely long periodic models. For an infinitely long periodic model, the eigenvalues of the transfer matrix relating the EM fields at $x$=$x_0$+$\Lambda$ to the EM fields at $x$=$x_0$ can be written as $\exp(ik_x\Lambda)$ in terms of the quasimomentum $k_x$ corresponding to $\omega$, where $x_0$ is an arbitrary point which does not coincide with the positions of the surface conductivities. The numerical values of $\varepsilon_b$ and $\varepsilon_m$ will hereafter be assumed to be equal to 1 and $(1.7)^2$, respectively. Also, we will always normalize $\omega$ and $\sigma_S$ to the Bragg angular frequency ($\omega_B$) and the free-space admittance ($Y_0$), respectively. It should be noted that, unlike the periodic model examined in the previous subsection, $\omega_B$ is now equal to $\pi c/(\sqrt{\varepsilon_m}\Lambda)$ [not $\pi c/(\sqrt{\varepsilon_b}\Lambda)$].

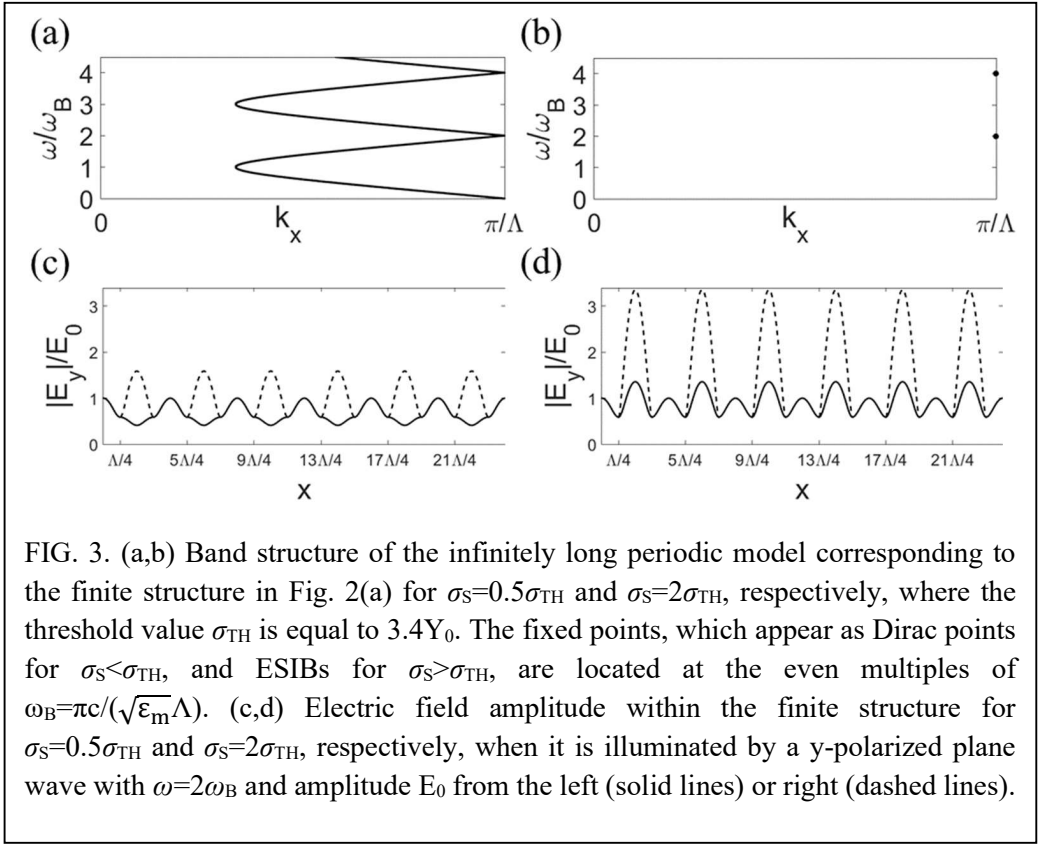

FIG. 3. (a,b) Band structure of the infinitely long periodic model corresponding to the finite structure in Fig. 2(a) for $\sigma_S=0.5\sigma_{TH}$ and $\sigma_S=2\sigma_{TH}$, respectively, where the threshold value $\sigma_{TH}$ is equal to $3.4Y_0$. The fixed points, which appear as Dirac points for $\sigma_S<\sigma_{TH}$, and ESIBs for $\sigma_S>\sigma_{TH}$, are located at the even multiples of $\omega_B=\pi c/(\sqrt{\varepsilon_m}\Lambda)$. (c,d) Electric field amplitude within the finite structure for $\sigma_S=0.5\sigma_{TH}$ and $\sigma_S=2\sigma_{TH}$, respectively, when it is illuminated by a y-polarized plane wave with $\omega=2\omega_B$ and amplitude $E_0$ from the left (solid lines) or right (dashed lines).

For $\sigma_S=0$, the EM structure in Fig. 2(a) reduces to a uniform medium of the relative permittivity $\varepsilon_m$. As is seen in Figs. 3(a,b), the band structure of the corresponding infinitely long periodic model has k-jumps for any non-zero value of $\sigma_S$. At any odd multiple of $\omega_B$, the Bloch wave propagating parallel to $+\hat{x}$ (or $-\hat{x}$) has a positive (or negative) k-jump from $-k_G$ to $k_G$ (or from $k_G$ to $-k_G$), where $0<k_G<\pi/\Lambda$. The k-jumps enlarge and approach $2\pi/\Lambda$ (viz., $k_G$ increases and approaches $\pi/\Lambda$) when $\sigma_S$ increases and approaches a threshold $\sigma_{TH}$ equal to $3.4Y_0$. For $\sigma_S>\sigma_{TH}$, a bandgap covering the whole frequency space, excluding all even multiples of $\omega_B$, appears. For $\sigma_S>\sigma_{TH}$, we name the states corresponding to the even multiples of $\omega_B$ 'extended states in the bandgap' (ESIB). These discrete points in the bandgap are in fact 'fixed points' of the band structure and appear as Dirac points for $\sigma_S<\sigma_{TH}$. The presence of the fixed points is an important feature which is absent in the band structure examined in Fig. 1.

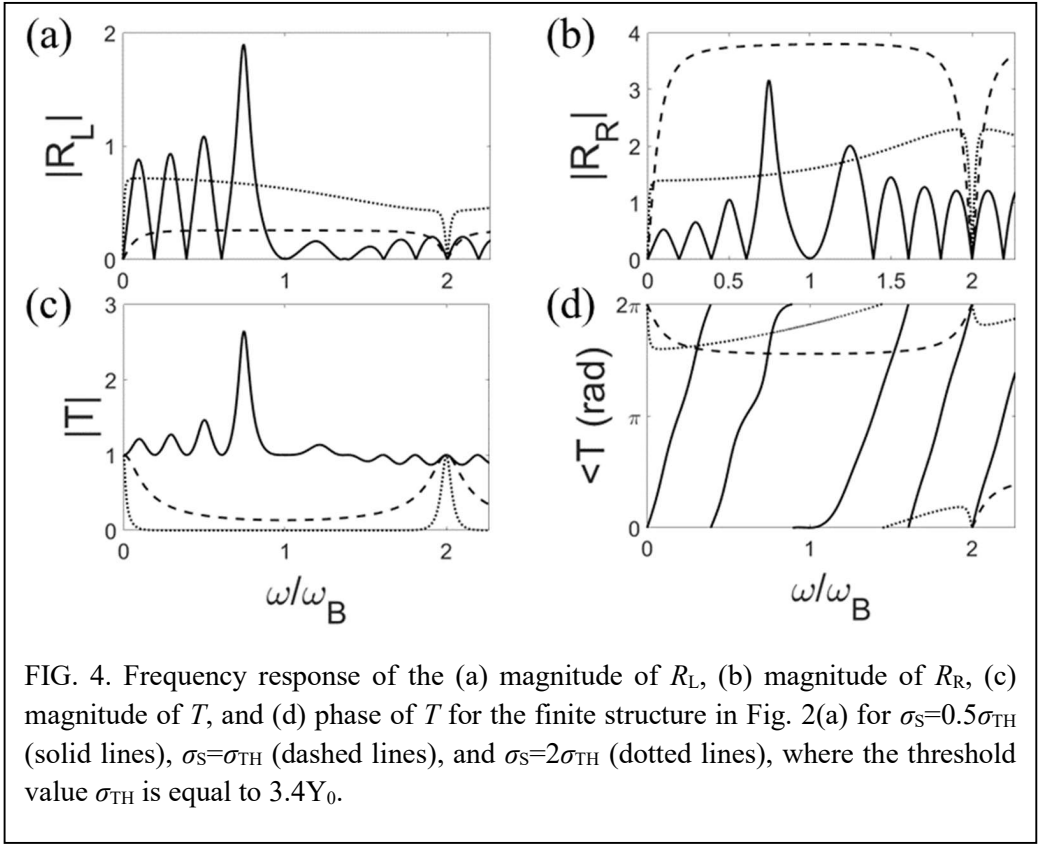

FIG. 4. Frequency response of the (a) magnitude of $R_L$, (b) magnitude of $R_R$, (c) magnitude of $T$, and (d) phase of $T$ for the finite structure in Fig. 2(a) for $\sigma_S=0.5\sigma_{TH}$ (solid lines), $\sigma_S=\sigma_{TH}$ (dashed lines), and $\sigma_S=2\sigma_{TH}$ (dotted lines), where the threshold value $\sigma_{TH}$ is equal to $3.4Y_0$.

As the fixed points of the band structure have quasimomentums equal to $\pi/\Lambda$, we expect the electric field amplitude within the finite structure depicted in Fig. 2(a) to be 'fully' periodic irrespective of the value of $\sigma_S$ when it is illuminated by a plane wave with an angular frequency equal to an even multiple of $\omega_B$. This prediction is confirmed by the numerical results presented in Figs. 3(c,d). More interestingly, the peak field amplitude within certain regions of the finite structure is fixed and equal to the incident field amplitude irrespective of the value of $\sigma_S$. Furthermore, as is seen in Figs. 4(a,b), the reflection coefficient is zero at the even multiples of $\omega_B$ irrespective of the value of $\sigma_S$, whether the structure is illuminated from the left or right. These 'bidirectional' reflection zeros contrast with the common belief that the reflection zeros of a structure with **PT** symmetry (and without **P** and **T** symmetries) are unidirectional [27].

The existence of the fixed points at the 'even' multiples of $\omega_B$ is not the sole interesting feature of the band structure. As is seen in Fig. 3(a), for $\sigma_S<\sigma_{TH}$, the group velocity $\partial\omega/\partial k_x$ of the Bloch wave propagating parallel to $+\hat{x}$ (or $-\hat{x}$) approaches $+\infty$ (or $-\infty$) when $\omega$ approaches any 'odd' multiple of $\omega_B$. Therefore, for $\sigma_S<\sigma_{TH}$, we expect a divergent light transmission velocity (viz., superluminal tunneling) in the finite structure with excitation frequencies around any 'odd' multiple of $\omega_B$. Figure 4(d), which depicts the phase $\phi$ of the transmission coefficient versus $\omega$, confirms that, for $\sigma_S<\sigma_{TH}$, $\phi$ is almost uniform around $\omega=\omega_B$, and therefore, the light transmission velocity, defined as $w(\partial\phi/\partial\omega)^{-1}$, diverges around $\omega=\omega_B$. The interesting point about this superluminal tunneling is that the bandwidth over which $\phi$ is almost uniform becomes very broad and almost equal to $\omega_B$ when $\sigma_S$ increases and approaches $\sigma_{TH}$.

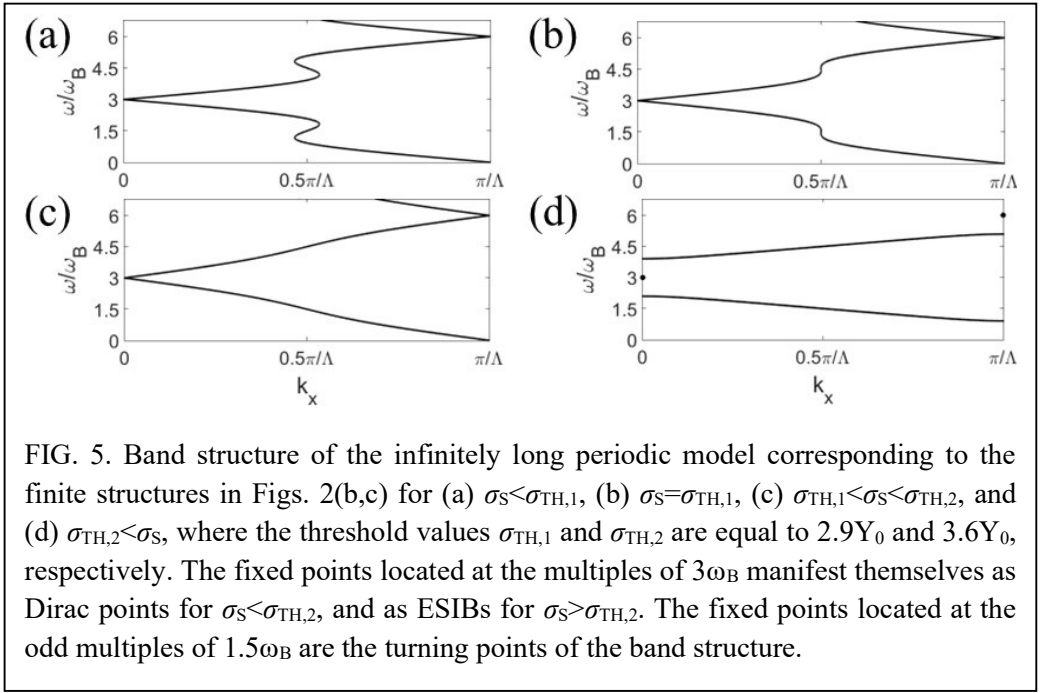

FIG. 5. Band structure of the infinitely long periodic model corresponding to the finite structures in Figs. 2(b,c) for (a) $\sigma_S<\sigma_{TH,1}$, (b) $\sigma_S=\sigma_{TH,1}$, (c) $\sigma_{TH,1}<\sigma_S<\sigma_{TH,2}$, and (d) $\sigma_{TH,2}<\sigma_S$, where the threshold values $\sigma_{TH,1}$ and $\sigma_{TH,2}$ are equal to $2.9Y_0$ and $3.6Y_0$, respectively. The fixed points located at the multiples of $3\omega_B$ manifest themselves as Dirac points for $\sigma_S<\sigma_{TH,2}$, and as ESIBs for $\sigma_S>\sigma_{TH,2}$. The fixed points located at the odd multiples of $1.5\omega_B$ are the turning points of the band structure.

Despite the obvious superiority of the superluminal tunneling discussed above over the schemes reported in the literature [28-32], two imperfections still make it non-ideal. First, when $\sigma_S$ is close to $\sigma_{TH}$ (and therefore the bandwidth of the superluminal tunneling is maximum), none of the reflection coefficients (viz., neither $R_L$ nor $R_R$) is close to zero around the odd multiples of $\omega_B$. Second, and more importantly, the magnitude of the transmission coefficient is not close to unity. These two imperfections can be explained based on the band structure of the periodic model. As $\sigma_S$ increases and approaches $\sigma_{TH}$, the states corresponding to the frequencies around any 'odd' multiple of $\omega_B$ approach 'localized' states in the band gap. Consequently, $R_L$ and $R_R$ depart from zero, and $|T|$ departs from unity when $\sigma_S$ increases and approaches $\sigma_{TH}$. As we will see, these two imperfections are absent in the EM structures depicted in Figs. 2(b,c).

The two finite structures in Figs. 2(b,c) correspond to one periodic model. The periodic model reduces to a uniform medium of the relative permittivity $\varepsilon_m$ when $\sigma_S=0$. The band structure of the periodic model has been plotted in Fig. 5 for different values of $\sigma_S$. For any non-zero value of $\sigma_S$ smaller than a threshold $\sigma_{TH,1}$ equal to $2.9Y_0$, the band structure has two types of k-jumps. In the case of any k-jump of the first type, the Bloch wave propagating parallel to $+\hat{x}$ (or $-\hat{x}$) has a positive (or negative) k-jump from $-k_A$ to $k_A$ (or from $k_A$ to $-k_A$), where $0<k_A<\pi/\Lambda$. In the case of any k-jump of the second type, the Bloch wave propagating parallel to $+\hat{x}$ (or $-\hat{x}$) has a positive (or negative) k-jump from $k_B$ to $2\pi/\Lambda-k_B$ (or from $2\pi/\Lambda-k_B$ to $k_B$), where $0<k_B<\pi/\Lambda$. The k-jumps all enlarge and approach $\pi/\Lambda$ (viz., $k_A$ and $k_B$ both increase and approach $0.5\pi/\Lambda$) when $\sigma_S$ increases and approaches $\sigma_{TH,1}$. However, at the same time, something interesting happens. For the sake of clarity, let us only consider the Bloch waves propagating either parallel to $+\hat{x}$ or $-\hat{x}$. For any k-jump of the first kind with a start point $S_A$ and an end point $E_A$ on the band structure, there is a k-jump of the second type with a start point $S_B$ and an end point $E_B$ with the property that $E_A$ and $E_B$ approach $S_B$ and $S_A$, respectively, when $\sigma_S$ increases and approaches $\sigma_{TH,1}$. As a result, the two k-jumps make a loop, and cancel out each other when $\sigma_S$ becomes equal to $\sigma_{TH,1}$. This is why the band structure plotted in Fig. 5 has no k-jumps when $\sigma_S$ is equal to or larger than $\sigma_{TH,1}$. This feature of the band structure contrasts with the common belief that some or all eigenvalues of a non-Hermitian operator become complex when non-Hermiticity becomes large enough.

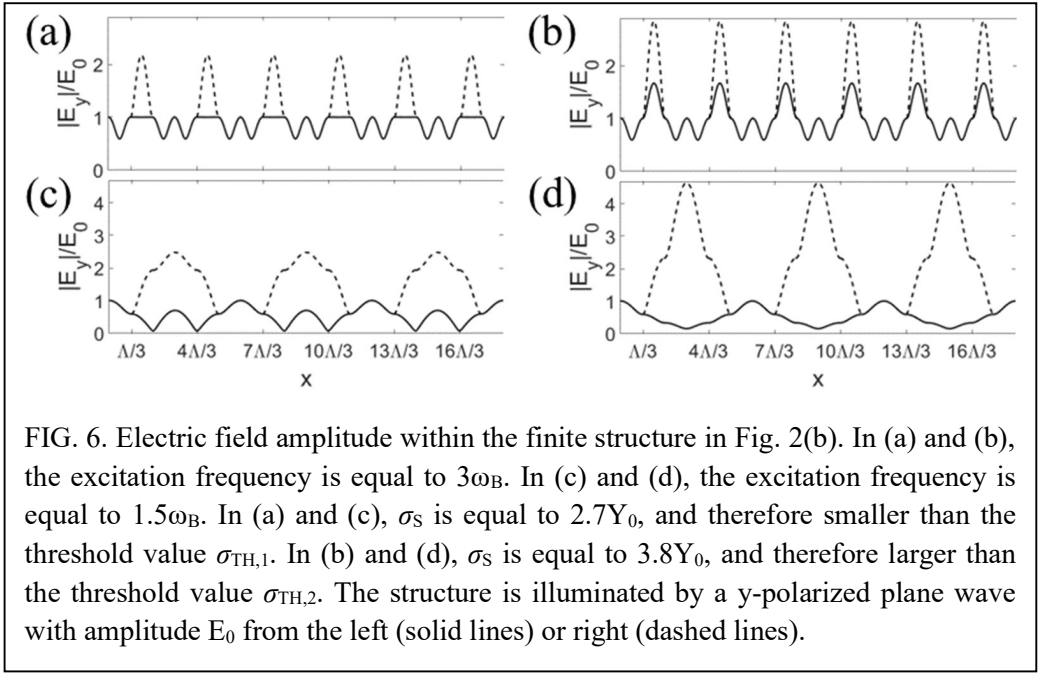

FIG. 6. Electric field amplitude within the finite structure in Fig. 2(b). In (a) and (b), the excitation frequency is equal to $3\omega_B$. In (c) and (d), the excitation frequency is equal to $1.5\omega_B$. In (a) and (c), $\sigma_S$ is equal to $2.7Y_0$, and therefore smaller than the threshold value $\sigma_{TH,1}$. In (b) and (d), $\sigma_S$ is equal to $3.8Y_0$, and therefore larger than the threshold value $\sigma_{TH,2}$. The structure is illuminated by a y-polarized plane wave with amplitude $E_0$ from the left (solid lines) or right (dashed lines).

The fixed points located at the multiples of $3\omega_B$ on the band structure in Fig. 5 manifest themselves as Dirac points when $\sigma_S$ is smaller than a threshold $\sigma_{TH,2}$ equal to $3.6Y_0$, and as ESIBs when $\sigma_S$ becomes larger than $\sigma_{TH,2}$. As is expected, and seen in Figs. 6(a,b), the electric field amplitude within the finite structure is 'fully' periodic irrespective of the value of $\sigma_S$ when it is illuminated by a plane wave at a multiple of $3\omega_B$. Also, the peak field amplitude within certain regions of the finite structure is fixed and equal to the incident field amplitude irrespective of the value of $\sigma_S$.

Apart from the fixed points corresponding to the multiples of $3\omega_B$ on the band structure, there exist other fixed points which are located at the odd multiples of $1.5\omega_B$ and are in fact the 'turning points' of the band structure. The existence of the turning points on the band structure in Fig. 5 is a feature which is absent in the band structure in Figs. 3(a,b). As the turning points have quasimomentums equal to $\pm 0.5\pi/\Lambda$, we expect the electric field amplitude within the finite structure to be 'fully' periodic irrespective of the value of $\sigma_S$ when it is illuminated by a plane wave with an angular frequency equal to an odd multiple of $1.5\omega_B$. This prediction is confirmed by the numerical results presented in Fig. 6(c,d).

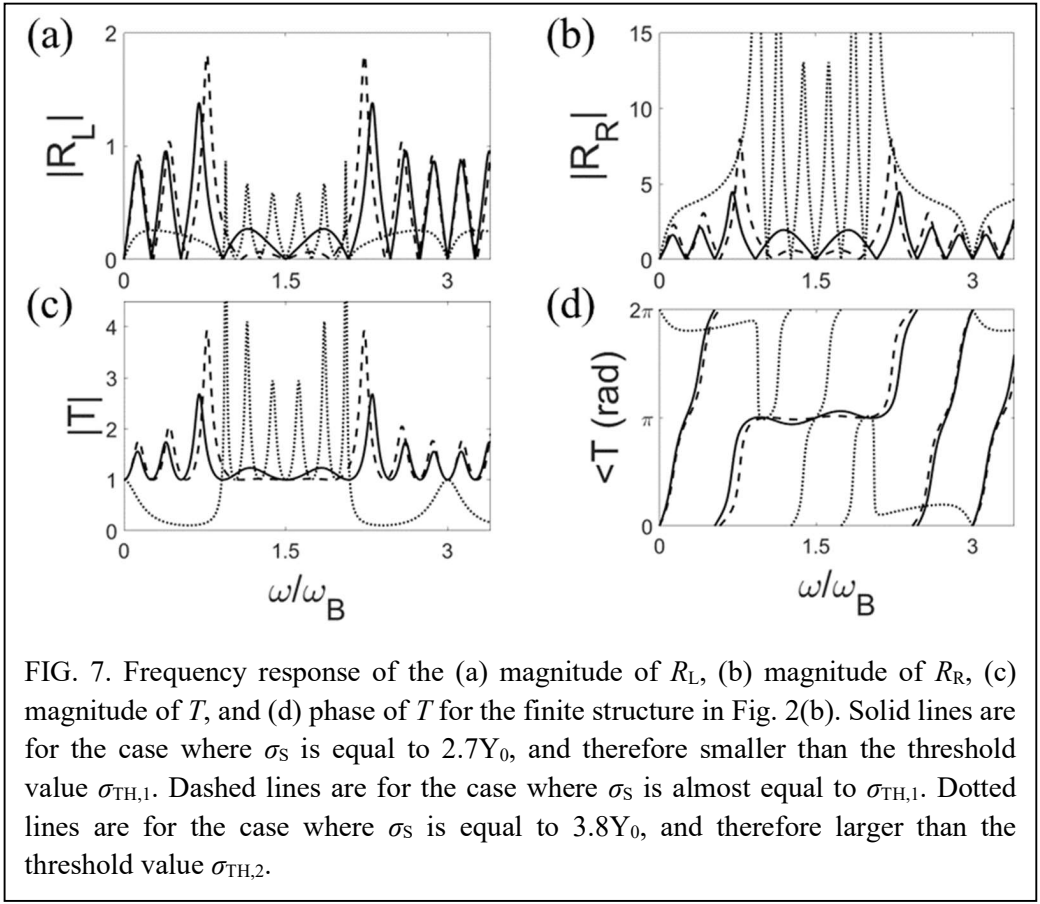

FIG. 7. Frequency response of the (a) magnitude of $R_L$, (b) magnitude of $R_R$, (c) magnitude of $T$, and (d) phase of $T$ for the finite structure in Fig. 2(b). Solid lines are for the case where $\sigma_S$ is equal to $2.7Y_0$, and therefore smaller than the threshold value $\sigma_{TH,1}$. Dashed lines are for the case where $\sigma_S$ is almost equal to $\sigma_{TH,1}$. Dotted lines are for the case where $\sigma_S$ is equal to $3.8Y_0$, and therefore larger than the threshold value $\sigma_{TH,2}$.

As is seen in Figs. 7(a,b), the reflection coefficient is zero at all the frequencies corresponding to the fixed points (viz., at all the multiples of $1.5\omega_B$) irrespective of the value of $\sigma_S$, whether the finite structure is illuminated from the left or right. These 'bidirectional' reflection zeros contrast with the common belief about the reflection zeros of **PT**-symmetric structures [27].

The turning points are especially interesting when $\sigma_S$ approaches $\sigma_{TH,1}$. At any odd multiple of $1.5\omega_B$, the group velocity of the Bloch wave propagating parallel to $+\hat{x}$ (or $-\hat{x}$) approaches $+\infty$ (or $-\infty$) when $\sigma_S$ approaches $\sigma_{TH,1}$. Therefore, we expect superluminal tunneling in the finite structure with excitation frequencies around any odd multiple of $1.5\omega_B$ when $\sigma_S$ approaches $\sigma_{TH,1}$. Figure 7(d) confirms that the phase of the transmission coefficient is almost uniform around $\omega=1.5\omega_B$ over a very broad bandwidth when $\sigma_S$ approaches $\sigma_{TH,1}$. This superluminal tunneling is superior to the one discussed above for the structure depicted in Fig. 2(a) for two reasons. First, as is seen in Figs. 7(a,b), the reflection coefficients (viz., $R_L$ and $R_R$) are both zero at the odd multiples of $1.5\omega_B$. Second, and more importantly, as is seen in Fig. 7(c), the magnitude of the transmission coefficient is equal to unity at the odd multiples of $1.5\omega_B$. We note that this superluminal tunneling is bidirectional because $R_L$ and $R_R$ are both zero at the odd multiples of $1.5\omega_B$. However, we probably prefer illuminating the structure from the left because $R_L$ remains small over a broad bandwidth around the odd multiples of $1.5\omega_B$.

Finally, we note that the value found for $\sigma_{TH,1}$ by using the band structure of the infinitely long periodic model is almost the same as the value found by using the phase of the transmission coefficient of the finite structure with six periods (i.e., M=6). However, if we reduce the length of the finite structure, we will see that the two values start to become slightly different.

### III. CONCLUDING REMARKS

We introduced the band-structure concept of 'fixed points' by examining one-dimensional **PT**-symmetric non-Hermitian EM models. The fixed points are the points on the band structure which remain intact when a certain parameter value of the model changes.

Two important EM phenomena occur in the finite structures corresponding to the periodic models when the excitation frequency is equal to the eigenfrequency of a fixed point. First, we observe 'bidirectional' reflection zeros, which contrast with the common belief that the reflection zeros of structures with **PT** symmetry (and without **P** and **T** symmetries) are unidirectional [27]. Second, the peak field amplitude within certain regions of the finite structures remains fixed and equal to the incident field amplitude, irrespective of the parameter value.

Some of the fixed points manifest themselves as what we name 'extended states in the bandgap' (ESIB), which can be considered as the dual of the well-known 'bound states in the continuum' [33-37]. Similar states in non-Hermitian systems have recently been reported in [38].

Some other fixed points are in fact the 'turning points' of the band structure. The turning points allow us to observe 'ideal' superluminal tunneling. It is noteworthy that superluminal tunneling [28-32,39,40] is in fact a combination of two phenomena which do not necessarily accompany each other: tunneling as the energy transmission through barriers [41], and superluminal behavior as unusually small group delays [28-30,42-45]. By 'ideal' superluminal tunneling we specifically refer to the case which is also accompanied by zero reflection and unit transmission over a broad bandwidth. In the cases reported in the literature [28-32], the reflection coefficient is not zero, and the magnitude of the transmission coefficient is usually very small.

The band-structure concepts introduced in this article might be useful in designing EM and electronic structures. Quantum-optical treatment of the phenomena reported in this article as well as the inclusion of nonlinear effects remains a subject for future work. Finding the electronic counterparts of the EM structures and phenomena examined here is another interesting problem.

## IV. ACKNOWLEDGMENTS

The author is very grateful to Prof. Aristide Dogariu for reviewing the last version of this article and his generous support over the past months. The author is also indebted to Prof. Behzad Rejaei for reviewing the first version of this article and having invaluable discussions with the author in the past years.